\documentclass{dcds}
\usepackage{amsmath, amsfonts, amssymb, amsthm}
\numberwithin{equation}{section}
 \usepackage[colorlinks=true]{hyperref}
\hypersetup{urlcolor=blue, citecolor=red}

\def\currentvolume{36}
 \def\currentissue{5}
  \def\currentyear{2016}
   \def\currentmonth{May}
    \def\ppages{X--XX}
     \def\DOI{10.3934/dcds.2016.36.xx}

\title[Persistence Properties and Unique Continuation for a 2CH]{Persistence Properties and Unique Continuation for a Dispersionless Two-Component
Camassa-Holm System with Peakon and Weak Kink Solutions}

\author[Qiaoyi Hu and Zhijun Qiao]{}


\email{huqiaoyi@scau.edu.cn}
\email{zhijun.qiao@utrgv.edu}

\begin{document}
\maketitle

\centerline{\scshape Qiaoyi Hu$^{1,2}$ and Zhijun Qiao$^2$}
\medskip
{\footnotesize
 \centerline{$^1$ Department of Mathematics}
   \centerline{South China
Agricultural University}
   \centerline{510642 Guangzhou, China}
   \centerline{$^2$ School of Mathematical and Statistical Sciences}
   \centerline{University of Texas - Rio Grande Valley}
   \centerline{78539 Edinburg, TX, USA}}

\bigskip

\centerline{(Communicated by Adrian Constantin)}

\begin{abstract}
In this paper, we study the persistence properties and unique continuation for a dispersionless two-component system with peakon and weak kink solutions. These properties guarantee 
strong solutions of the two-compon\-ent system decay at infinity in the
spatial variable provided that the initial data satisfies the condition of decaying at infinity.
Furthermore, we give an optimal decaying index of the momentum for the system and show 
that the system exhibits unique continuation if the initial momentum $m_0$ and $n_0$ are non-negative.
\end{abstract}

\section{Introduction}
Recently, an integrable two-component Camassa-Holm
system with both quadratic and cubic nonlinearity was proposed by Xia and Qiao \cite{Xia1}
\begin{equation}
\left\{\begin{array}{ll}m_{t}+\frac{1}{2}[m(uv-u_xv_x)]_x-\frac{1}{2}m(uv_x-u_xv)+bu_x=0 \\
n_{t}+\frac{1}{2}[n(uv-u_xv_x)]_x+\frac{1}{2}n(uv_x-u_xv)+bv_x=0\\
m=u-u_{xx},~n=v-v_{xx}.
\end{array}\right.
\end{equation} As shown in \cite{Xia1}, this 
system has peakon and weak kink solutions 
as well as including some remarkable peakon equations such as the CH equation and the FORQ equation.
For instance, letting $v = 2$ in Eq.(1.1) yields the Camassa-Holm (CH) equation,
which models the unidirectional propagation of shallow water waves
over a flat bottom while $u(t,x)$ stands for the fluid velocity at
time $t$ in the spatial $x$ direction\cite{C-H,C-L,J}. The CH equation has a bi-Hamiltonian structure \cite{C1,
F-F} and is completely integrable \cite{C-H,C-Mc, C3}. 
The Cauchy problem of the CH equation has been studied
extensively. This equation is locally
well-posed \cite{C4,C-E2,L-O, Rb} for initial data $u_{0}\in
H^{s}(\mathbb{S})$ with $s>\frac{3}{2}$. More interestingly, it
has not only global strong solutions modelling permanent waves
\cite{C-E2} and but also blow-up solutions modelling
wave breaking \cite{C2,C-E1, C-E2, C-E3,L-O,Rb}. On the
other hand, it has globally weak solutions with initial data $u_0\in
H^1$, cf. \cite{B-C,C-M,X-Z}.

If choosing $v =2u$ in Eq. (1.1), one may obtain the cubic CH equation which is also called the FORQ equation in the literature since it was developed independently in \cite{F-F,Olver,Qiao1,Qiao2}. It might be derived from the two dimensional
Euler equations, and its Lax pair, cuspon and other peaked solutions
have been studied in \cite{Qiao1,Qiao2}. 

With $v = k_1u +k_2,$ Eq. (1.1) is able to be reduced to the generalized CH (gCH) equation.
The gCH equation was first implied in the work of Fokas \cite{Fokas2}. Its Lax pair, bi-Hamiltonian structure, peakons, weak kinks, kink-peakon interaction, and
classical soliton solutions were investigated in \cite{Q-X}.

Moreover, by imposing the constraint 
$v =u*$, equation (1.1) is reduced to a new
integrable equation with cubic nonlinearity and linear dispersion
\begin{equation}
m_t =bu_x + \frac{1}{2}
[m(|u|^2-|u_x|^2)]_x -\frac{1}{2}
m(uu^*_x -u_xu^*), m=u- u_{xx}
\end{equation}
where the symbol $*$ denotes the complex conjugate.
The above reduction of the two-component system (1.1) looks very like the reduction case of AKNS system, which embraces the KdV
equation, the mKdV equation, the Gardner equation, and the nonlinear Schr\"{o}dinger equation. Xia and Qiao \cite{Xia1,Xia2} proposed the complex-value N-peakon
solution and weak kink wave solution to the cubic nonlinear equation (1.2).

Geometrically, system (1.1) describes pseudo-spherical surfaces.
Integrability of the system, its bi-Hamiltonian structure, and infinitely many conservation laws were already presented by Xia and Qiao \cite{Xia1}. In the case $b=0$ (dispersionless
case), the authors showed that this system admits the single-peakon of travelling wave solution as well
as multi-peakon solutions. The qualitative analysis for the integrable system (1.1) was investigated by Yan, Qiao and Yin \cite{Yan}.

In this paper, we consider the following Cauchy problem of
system (1.1) with $b=0$ on the line:
\begin{equation}
\left\{\begin{array}{ll}m_{t}+\frac{1}{2}[m(uv-u_xv_x)]_x-\frac{1}{2}m(uv_x-u_xv)=0 ,&t > 0,\,x\in\mathbb{R}, \\
n_{t}+\frac{1}{2}[n(uv-u_xv_x)]_x+\frac{1}{2}n(uv_x-u_xv)=0, &t>0, x\in\mathbb{R},\\
m(0,x)=m_0,~n(0,x)=n_0, &x\in\mathbb{R} ,
\end{array}\right.
\end{equation}
where $m=u-u_{xx}$ and $n=v-v_{xx}$, and
study the persistence properties and unique continuation of strong solutions for Eq.(1.3). There is a lot literatures concerning these problems. The persistence properties and unique continuation for the CH equation are proved in \cite{Him}. The unique continuation results about the Schr\"{o}dinger and KdV equations were provided by Escauriaza, Kenig, Ponce and Vega in \cite{EKPV1} and \cite{EKPV2}. Persistence properties and infinite propagation for the modified 2-component Camassa-Holm equation and 3-component Camassa-Holm system were investigated in \cite{Henry,Wu1}. 

As we mentioned at the very beginning of the paper, 
system (1.1) possesses peakons and weak kink solutions with both quadratic and cubic nonlinearity.
It is quite interesting to study the persistence properties and unique continuation of strong solutions for system (1.1).
Inspired by the method given by Himonas et al. in \cite{Him},
we will show some persistence properties of the strong solutions, and furthermore present the optimal decay index of the momentum. Finally, by introducing a continuous family of diffeomorphisms of the line, we demonstrate that the system
exhibits unique continuation if the initial momentum $m_0$ and $n_0$ are non-negative.

\medskip
\par
\noindent \textbf{Notation.} Throughout this paper, the convolution is denoted by
$\ast$. For
$ 1 \leq p \leq \infty, $ the norm in the Lebesgue space $
L^p(\mathbb{R})$ is written by $ \| \cdot \|_{L^p}, $ while $
\| \cdot \|_{H^s}, \, s
> 0$, stands for the norm in the classical Sobolev spaces $
H^s(\mathbb{R}).$

\section{Persistence properties}
\newtheorem{theorem2}{Theorem}[section]
\newtheorem{lemma2}{Lemma}[section]
{\theoremstyle{definition}\newtheorem{remark2}{Remark}[section]
\newtheorem{definition2}{Definition}[section]}
\par
For our convenience, let us first present the following well-posedness theorem given in \cite{Yan}.
\begin{theorem2}\cite{Yan}
Let $s \geq 3$. If $z_0=(u_0,v_0)$ belongs to the Sobolev space $H^s\times H^s$ on the circle or the line, then
there exists a maximal time $T = T(z_0) > 0$ and a unique solution $z(t,x) \in C([0; T);H^s\times H^s)\cap C^1([0; T);H^{s-1}\times H^{s-1})$ of the Cauchy problem for the
equation (1.1). Furthermore, the data-to-solution map $z(0) \rightarrow z(t)$ is
continuous but not uniformly continuous.
\end{theorem2}
From the above well-posedness result, we may now utilize it to the persistence properties and unique
continuation to equation (1.3). Our basic assumption is that the initial data and its first spacial derivative
decay exponentially. Then we have the following result based on the work \cite{Him} for the CH equation.
\begin{theorem2}
Assume that $s \geq 3, ~T > 0,$ and $z \in C([0; T];H^s\times H^s)$ is a solution of (1.3). If the
initial data $z_0(x) = z(0, x)$ decays at infinity, more precisely, if there is some $\theta \in (0, 1)$ such that
as $|x| \rightarrow \infty$
\begin{align*}
&|u_0(x)|\thicksim O(e^{-\theta |x|}), ~~ |u_0'(x)|\thicksim O(e^{-\theta |x|}),\\
&|v_0(x)|\thicksim O(e^{-\theta |x|}), ~~ |v_0'(x)|\thicksim O(e^{-\theta |x|})\end{align*}
then as $|x| \rightarrow \infty$, we have
\begin{align*}
&|u(t,x)|\thicksim O(e^{-\theta |x|}), ~~ |\partial_x u(t,x)|\thicksim O(e^{-\theta |x|}),\\
&|v(t,x)|\thicksim O(e^{-\theta |x|}), ~~ |\partial_x v(t,x)|\thicksim O(e^{-\theta |x|})
\end{align*}
uniformly with respect to $t \in [0, T].$
\end{theorem2}

After establishing unique continuation for system (1.3) in the sense of Theorem 2.2, it is natural
to ask the question of how the solution behaves at infinity when given compactly supported initial
data. This qualitative behavior is examined by Theorem 4.1.

The paper is organized as follows. In Section 3, we prove the persistence properties of system (1.3) as listed in Theorem 2.2. Then we prove the optimal decay index of the momentum $m$ and $n$. In Section 4 we examine the behavior of strong solutions when the initial data have compact support.

\section{Proof of Theorem 2.2}
\newtheorem{theorem3}{Theorem}[section]
\newtheorem{lemma3}{Lemma}[section]\newtheorem{corollary3}{Corollary}[section]
{\theoremstyle{definition}\newtheorem {remark3}{Remark}[section]}
\par
In the section, we prove the persistence properties of system (1.3). For our convenience, we rewrite Eq.(1.3) as the form of a quasi-linear evolution equation
of hyperbolic type. Note that $G(x):=\frac{1}{2}e^{-|x|}$ is the
kernel of $(1-
\partial^{2}_{x})^{-1}$. Then $(1-
\partial^{2}_{x})^{-1}f = G*f $ for all $f \in L^{2}(\mathbb{R})$, $G \ast
m=u $ and $G \ast n=v$. By these
identities, Eq.(1.3) can be reformulated as follows:
\begin{equation}
\left\{\begin{array}{ll} u_{t}+\frac{1}{2}(uv-u_xv_x)u_{x}=G\ast F_1+\partial_xG\ast F_2,&t
> 0,\,x\in\mathbb{R},\\
 v_{t}+\frac{1}{2}(uv-u_xv_x)v_{x}=G\ast H_1+\partial_xG\ast H_2,&t
> 0,\,x\in\mathbb{R},
\\ u(0,x) = u_{0}(x),&x\in
\mathbb{R},\\
v(0,x) =
v_{0}(x),&x\in\mathbb{R},\end{array}\right.
\end{equation}
where
\begin{equation*}M:=(u_x n+v_x m)=(uv-u_xv_x)_x,\end{equation*}
\begin{equation*}F_1:=-\frac{1}{2}(uM-(uv_x-u_xv)m),~~F_2:=-\frac{1}{2}(u_xM),\end{equation*}
 \begin{equation*}H_1:=-\frac{1}{2}(vM+(uv_x-u_xv)n),~~H_2:=-\frac{1}{2}(v_xM).\end{equation*}

 Assume that $z \in C([0; T];H^s\times H^s)$ is a strong solution to (1.3) with $s \geq 3.$ Let
$$
K = \sup_{t\in[0,T ]}\|z(t)\|_{H^s}:=\sup_{t\in[0,T ]}(\|u(t)\|_{H^s}+\|v(t)\|_{H^s}),
$$
hence by Sobolev imbedding theorem, we have
\begin{align}
& \|u(t,\cdot)\|_{L^{\infty}}+\|u_x(t,\cdot)\|_{L^{\infty}}+\|u_{xx}(t,\cdot)\|_{L^{\infty}}\leq CK,\\
&\|v(t,\cdot)\|_{L^{\infty}}+\|v_x(t,\cdot)\|_{L^{\infty}}+\|v_{xx}(t,\cdot)\|_{L^{\infty}}\leq CK
\end{align}
Set
\begin{equation}
\varphi _N(x)= \left\{\begin{array}{ll}e^{\theta |x|},~~|x|<N,\\
e^{N |x|},~|x|\geq N,\end{array}\right.
\end{equation}
where $N \in \mathbb{N}$ and $\theta \in (0, 1).$ Observe that for all $\mathbb{N}$ we have
\begin{equation}
0\leq |\varphi' _N|\leq \varphi _N(x),~~a.e.~~x\in \mathbb{R}.
\end{equation}

Multiplying (3.1)$_1$ by $(u\varphi _N)^{2q-1}\varphi _N$ for $q\in\mathbb{N}$ and integrating over the real line we obtain
\begin{align}
&\nonumber \frac{1}{2q}\frac{1}{dt}\int (u\varphi _N)^{2q}dx=-\frac{1}{2}\int (uv-u_xv_x)u_x(u\varphi _N)^{2q-1}\varphi _Ndx\\
&\phantom{ \frac{1}{2q}}+\int \partial_x(G\ast F_2)(u\varphi _N)^{2q-1}\varphi _N dx+\int (G\ast F_1)(u\varphi _N)^{2q-1}\varphi _N dx.
\end{align}

(3.2)-(3.3) and H\"{o}lder's inequality lead us to achieve the following estimates
\begin{equation}
|-\frac{1}{2}\int (uv-u_xv_x)u_x(u\varphi _N)^{2q-1}\varphi _Ndx|\leq CK^2 \|u\varphi _N\|^{2q-1}_{2q}\|u_x\varphi _N\|_{2q},
\end{equation}
\begin{equation}
|\int (G\ast F_1)(u\varphi _N)^{2q-1}\varphi _N dx|\leq \|u\varphi _N\|_{2q}^{2q-1}\|(G\ast F_1)\varphi _N\|_{2q},
\end{equation}
and
\begin{equation}
|\int \partial_x(G\ast F_2)(u\varphi _N)^{2q-1}\varphi _N dx|\leq \|u\varphi _N\|_{2q}^{2q-1}\|(\partial_xG\ast F_2)\varphi _N\|_{2q}.
\end{equation}

From (3.6) and the above estimates, this implies
\begin{equation}
\frac{d}{dt}\|u\varphi _N\|_{2q}\leq CK^2\|u\varphi _N\|_{2q}+\|(G\ast F_1)\varphi _N\|_{2q}+\|(\partial_xG\ast F_2)\varphi _N\|_{2q}.
\end{equation}
By Gronwall's inequality, (3.10) implies the following estimate
\begin{equation}
\|u\varphi _N\|_{2q}\leq (\|u_0\varphi _N\|_{2q}+\int_0^t[\|(G\ast F_1)\varphi _N\\|_{2q}+\|(\partial_xG\ast F_2)\varphi _N\|_{2q}]d\tau)e^{CK^2t}.
\end{equation}
Now differentiating (3.1)$_1$ with respect to the spacial variable $x$, multiplying by $(u_x\varphi _N)^{2q-1}\varphi _N$ and
integrating over the real line yields
\begin{align}
&\nonumber \frac{1}{2q}\frac{1}{dt}\int (u_x\varphi _N)^{2q}dx=-\frac{1}{2}\int (uv-u_xv_x)u_{xx}(u_x\varphi _N)^{2q-1}\varphi _Ndx\\
&\nonumber\phantom{ \frac{1}{2q}}+\int \partial_x^2(G\ast F_2)(u_x\varphi _N)^{2q-1}\varphi _N dx+\int \partial_x(G\ast F_1)(u_x\varphi _N)^{2q-1}\varphi _N dx\\
&-\frac{1}{2}\int M u_x(u_x\varphi _N)^{2q-1}\varphi _N dx.
\end{align}
This leads us to obtain the following estimates
\begin{equation*}
|\int \partial_x^2(G\ast F_2)(u_x\varphi _N)^{2q-1}\varphi _N dx|\leq \|u_x\varphi _N\|_{2q}^{2q-1}\|(\partial_x^2G\ast F_2)\varphi _N\|_{2q},
\end{equation*}
\begin{equation*}
|\int (\partial_x G\ast F_1)(u_x\varphi _N)^{2q-1}\varphi _N dx|\leq \|u_x\varphi _N\|_{2q}^{2q-1}\|(\partial_x G\ast F_1)\varphi _N\|_{2q},
\end{equation*}
\begin{equation}
|-\frac{1}{2}\int M u_x(u_x\varphi _N)^{2q-1}\varphi _N dx|\leq \|M\|_{L^{\infty}}\|u_x\varphi _N\|^{2q}_{2q}\leq CK^2 \|u_x\varphi _N\|^{2q}_{2q},
\end{equation}
For the first integral on the RHS of (3.12), we estimate as follows
\begin{align}
\nonumber&\int (uv-u_xv_x) u_{xx}(u_x\varphi _N)^{2q-1}\varphi _Ndx\\
\nonumber&=\int (uv-u_xv_x) [(u_x\varphi _N)_x -u_x\varphi _N'](u_x\varphi _N)^{2q-1}dx\\
\nonumber&=-\frac{1}{2q}\int M (u_x\varphi _N)^{2q}dx -\int (uv-u_xv_x)u_x\varphi _N'(u_x\varphi _N)^{2q-1}dx\\
&\leq CK^2\|u_x\varphi _N\|_{2q}^{2q}.
\end{align}
From (3.12) - (3.14), we achieve the following differential inequality
\begin{equation}
\frac{d}{dt}\|u_x\varphi _N\|_{2q}\leq CK^2\|u_x\varphi _N\|_{2q}+\|(\partial_x^2G\ast F_2)\varphi _N\|_{2q}+\|(\partial_xG\ast F_1)\varphi _N\|_{2q}.
\end{equation}
By Gronwall's inequality, (3.15) implies the following estimate
\begin{equation}
\|u_x\varphi _N\|_{2q}\leq (\|\partial_xu_0\varphi _N\|_{2q}+\int_0^t[\|(\partial_x^2G\ast F_2)\varphi _N\|_{2q}+\|(\partial_x G\ast F_1)\varphi _N\|_{2q}]d\tau)e^{CK^2t}.
\end{equation}
By adding (3.11) and (3.16), we have the following
\begin{align}
\nonumber&\|u\varphi _N\|_{2q}+\|u_x\varphi _N\|_{2q}\leq (\|u_0\varphi _N\|_{2q}+\|\partial_x u_0\varphi _N\|_{2q})e^{CK^2t}\\
\nonumber&+(\int_0^t[\|(\partial_xG\ast F_2)\varphi _N\|_{2q}+\|(G\ast F_1)\varphi _N\|_{2q}]d\tau)e^{CK^2t}\\
&+(\int_0^t[\|(\partial_x^2G\ast F_2)\varphi _N\|_{2q}+\|(\partial_x G\ast F_1)\varphi _N\|_{2q}]d\tau)e^{CK^2t}.
\end{align}
Now, for any function $f\in L^1\cap L^{\infty}$,$\lim_{n\rightarrow \infty} \|f\|_{L^n}=\|f\|_{L^{\infty}}.$ Since we have that $F_1,F_2 \in L^1\cap L^{\infty}$
and $G\in W^{1,1}$, we know that $\partial_x^i G \ast F_1,~\partial_x^j G \ast F_2\in L^1\cap L^{\infty}$
(for $i=0,1$ and $j=1,2$). Thus,
by taking the limit of (3.17) as $q\rightarrow \infty$, we get
\begin{align}
\nonumber&\|u\varphi _N\|_{_{\infty}}+\|u_x\varphi _N\|_{_{\infty}}\leq (\|u_0\varphi _N\|_{_{\infty}}+\|\partial_xu_0\varphi _N\|_{_{\infty}})e^{CK^2t}\\
\nonumber&+(\int_0^t[\|(\partial_xG\ast F_2)\varphi _N\|_{_{\infty}}+\|(G\ast F_1)\varphi _N\|_{_{\infty}}]d\tau)e^{CK^2t}\\
&+(\int_0^t[\|(\partial_x^2G\ast F_2)\varphi _N\|_{_{\infty}}+\|(\partial_x G\ast F_1)\varphi _N\|_{_{\infty}}]d\tau)e^{CK^2t}.
\end{align}
A simple calculation shows that for $\theta \in (0,1)$
\begin{equation}
\varphi_N(x)\int_{\mathbb{R}}e^{-|x-y|}\frac{1}{\varphi_N(y)}dy\leq \frac{4}{1-\theta}=C_0.
\end{equation}
Thus, for any function $f,g,h\in L^{\infty}$, we have
\begin{align*}
&\|(G\ast fgh)\varphi _N\|_{\infty}= \frac{1}{2}\varphi_N \int_{\mathbb{R}}e^{-|x-y|}(fgh)(y)dy\\
&\leq \frac{1}{2}(\varphi _N\int_{\mathbb{R}}e^{-|x-y|}\frac{1}{\varphi _N(y)}dy)\|f\|_{\infty}\|g\|_{\infty}\|h\varphi _N\|_{\infty}\\
&\leq C_0\|f\|_{\infty}\|g\|_{\infty}\|h\varphi _N\|_{\infty}.
\end{align*}
Similary, we have
\begin{align*}
&\|(\partial_xG\ast fgh)\varphi _N\|_{\infty}= \frac{1}{2}\varphi_N \int_{\mathbb{R}}e^{-|x-y|}(fgh)(y)dy\\
&\leq \frac{1}{2}(\varphi _N\int_{\mathbb{R}}e^{-|x-y|}\frac{1}{\varphi _N(y)}dy)\|f\|_{\infty}\|g\|_{\infty}\|h\varphi _N\|_{\infty}\\
&\leq C_0\|f\|_{\infty}\|g\|_{\infty}\|h\varphi _N\|_{\infty}.
\end{align*}
Therefore, since $u,v,u_x,v_x,m,n,M\in L^{\infty}$, we get
\begin{align*}
&\|(\partial_x^jG\ast uM)\varphi _N\|_{\infty}\leq C_0\|M\|_{\infty}\|u\varphi _N\|_{\infty}\leq C_0 K^2\|u\varphi _N\|_{\infty},~~j=0,1
\end{align*}
\begin{align*}
&~~~~\|(\partial_x^jG\ast (uv_xm-u_x vm)\varphi _N\|_{\infty}\\
&\leq C_0(\|v_xm\|_{\infty}\|u\varphi _N\|_{\infty}+\|vm\|_{\infty}\|u_x\varphi _N\|_{\infty})\\
&\leq C_0K^2(\|u\varphi _N\|_{\infty}+\|u_x\varphi _N\|_{\infty}),~~j=0,1
\end{align*}
hence,
\begin{align}
&\|(\partial_x^jG\ast F_1)\varphi _N\|_{\infty}\leq C_0K^2(\|u\varphi _N\|_{\infty}+\|u_x\varphi _N\|_{\infty})~~j=0,1.
\end{align}
Similarly, we have
\begin{align*}
&~~~~\|(\partial_x^jG\ast u_xM)\varphi _N\|_{\infty}
\leq C_0(\|M\|_{\infty}\|u_x\varphi _N\|_{\infty}
\leq C_0K^2\|u_x\varphi _N\|_{\infty},~~j=0,1
\end{align*}
For $j=2$, noticing that $\partial_x^2 G\ast f=G\ast f-f$, using the similar procedure, we have
\begin{align}
&\|(\partial_x^2G\ast u_xM)\varphi _N\|_{\infty}\leq C_0K^2\|u_x\varphi _N\|_{\infty}.
\end{align}
Thus, we obtain
\begin{align}
&\|(\partial_x^jG\ast F_2)\varphi _N\|_{\infty}\leq C_0K^2 \|u_x\varphi _N\|_{\infty}~~j=1,2.
\end{align}
So, by estimates (3.18), (3.20) and (3.22) we achieve the following
\begin{align}
\nonumber&\|u\varphi _N\|_{\infty}+\|u_x\varphi _N\|_{\infty}
\leq C(\|u_0\varphi _N\|_{\infty}+\|u_{0,x}\varphi _N\|_{\infty})\\
&+C\int_0^t(\|u_0\varphi _N\|_{\infty}+\|u_{0,x}\varphi _N\|_{\infty})d\tau
\end{align}
where $C$ is a constant depending on $C_0,K $ and $T$.

Multiplying (3.1)$_2$ by $(v\varphi _N)^{2q-1}\varphi _N$ for $q\in\mathbb{N}$ and integrating over the real line, then differentiating (3.1)$_2$ with respect to the spacial variable $x$, multiplying by $(v_x\varphi _N)^{2q-1}\varphi _N$ and integrating over the real line yields, using the similar steps above, we get
\begin{align}
\nonumber&\|v\varphi _N\|_{\infty}+\|v_x\varphi _N\|_{\infty}
\leq C(\|v_0\varphi _N\|_{\infty}+\|v_{0,x}\varphi _N\|_{\infty})\\
&+C\int_0^t(\|v_0\varphi _N\|_{\infty}+\|v_{0,x}\varphi _N\|_{\infty})d\tau
\end{align}
Adding (3.23) and (3.24), we have
\begin{align*}
&\|u\varphi _N\|_{\infty}+\|u_x\varphi _N\|_{\infty}+\|v\varphi _N\|_{\infty}\|v_x\varphi _N\|_{\infty}\\
&\leq
C\|u_0\varphi _N\|_{\infty}+\|v_0\varphi _N\|_{\infty}+\|u_{0,x}\varphi _N\|_{\infty}+\|v_{0,x}\varphi _N\|_{\infty}\\
&+C \int_0^t(\|u\varphi _N\|_{\infty}+\|v\varphi _N\|_{\infty}+\|u_x\varphi _N\|_{\infty}+\|v_x\varphi _N\|_{\infty})d\tau.
\end{align*}
Hence, for any $N\in\mathbb{N}$ and any $t\in [0,T]$, we have by Gronwall's inequality that
\begin{align}
\nonumber&\|u\varphi _N\|_{\infty}+\|u_x\varphi _N\|_{\infty}+\|v\varphi _N\|_{\infty}+\|v_x\varphi _N\|_{\infty}\\
\nonumber&\leq
C(\|u_0\varphi _N\|_{\infty}+\|v_0\varphi _N\|_{\infty}+\|u_{0,x}\varphi _N\|_{\infty}+\|v_{0,x}\varphi _N\|_{\infty})\\
&\leq C(\|u_0 f_{\theta})\|_{\infty}+\|v_0f_{\theta})\|_{\infty}+\|u_{0,x}f_{\theta})\|_{\infty}
+\|v_{0,x}f_{\theta}\|_{\infty}),
\end{align}
with $f_{\theta}:=\max(1,e^{\theta |x|}).$
This concludes our proof of Theorem 2.2.
\begin{remark3}
In fact, let $\theta\in(0,1)$, and $j=0,1,2,...,$ if the initial data $z_0$ satisfy
$$
\partial_x^j u_{0},\partial_x^j v_{0}\thicksim O(e^{-\theta|x|}),~as~|x|\rightarrow \infty,
$$
then the solution $z$ also has the same exponential decay properties, i.e.
$$
\partial_x^j u,\partial_x^j v\thicksim O(e^{-\theta|x|}),~as~|x|\rightarrow \infty.
$$
\end{remark3}
Theorem 2.2 tells us that the solution $z$ can decay as
$e^{-\theta|x|},~as~x\rightarrow \infty$ for $\theta\in(0,1)$. Whether the decay is
optimal? the next result tell us some information.
\begin{theorem3}
Given $z_0 = (u_0, v_0) \in H^s \times H^s, s \geq 3.$ Let $T =
T(z_0)$ be the maximal existence time of the solutions $z(t,x) = (u(t,x), v(t,x))$ to
system (1.3)(or (3.1)) with the initial data $z_0.$ If for some $\lambda \geq 0$ and $q \geq 1,$
\begin{equation}
\|(m_0,n_0)e^{(1+\lambda)|x|}\|_{L^{2q}}\leq C,
\end{equation}
then for all $t\in[0,T)$,  we have
\begin{equation}
\|(m,n)e^{(1+\lambda)|x|}\|_{L^{2q}}\leq C,
\end{equation}
Moreover, if the initial data satisfy
\begin{equation}
\partial_x^ju_{0},\partial_x^jv_{0}\thicksim O(e^{-(1+\lambda)|x|}),~as~|x|\rightarrow \infty,j=0,1,2,
\end{equation}
then for all $t\in[0,T)$,  we get
\begin{equation}
m,n\thicksim O(e^{-(1+\lambda)|x|}),~as~|x|\rightarrow \infty,
\end{equation}
and there exists some $\theta\in (0, 1)$ such that
\begin{equation}
\lim_{|x|\rightarrow \infty}|(\partial_x^ju,\partial_x^jv)e^{-\theta|x|}|\leq C,j=0,1,2.
\end{equation}
\end{theorem3}
\begin{proof}
Setting $\varphi _{\lambda}:= e^{(1+\lambda)|x|}$, multiplying (3.1)$_1$ by $(m\varphi _{\lambda})^{2q-1}\varphi _{\lambda}$ for $q\in\mathbb{N}$ and integrating over the real line we obtain
\begin{align}
&\nonumber \frac{1}{2q}\frac{1}{dt}\int (m\varphi _{\lambda})^{2q}dx=-\frac{1}{2}\int (uv-u_xv_x)m_x(m\varphi _{\lambda})^{2q-1}\varphi _{\lambda}dx\\
&\phantom{ \frac{1}{2q}}-\frac{1}{2}\int M m(m\varphi _{\lambda})^{2q-1}\varphi _{\lambda} dx+\frac{1}{2}\int (uv_x-u_x v)(m\varphi _{\lambda})^{2q-1}m\varphi _{\lambda} dx.
\end{align}
For the first term on RHS of (3.31), we have
\begin{align*}
\nonumber&\int (uv-u_xv_x) m_{x}(m\varphi _{\lambda})^{2q-1}\varphi _{\lambda} dx\\
\nonumber&=\int (uv-u_xv_x) [(m\varphi _{\lambda})_x -m\varphi _{\lambda}'](m\varphi _{\lambda})^{2q-1}dx\\
\nonumber&=-\frac{1}{2q}\int M (m\varphi _{\lambda})^{2q}dx -\int (uv-u_xv_x)m\varphi _{\lambda}'(m\varphi _{\lambda})^{2q-1}dx\\
&= -\frac{1}{2q}\int M (m\varphi _{\lambda})^{2q}dx -(1+\lambda)\int sgn(x) (uv-u_xv_x)m\varphi _{\lambda}(m\varphi _{\lambda})^{2q-1}dx,
\end{align*}
where we use the fact that \begin{equation*}\varphi _{\lambda}'=(1+\lambda)sgn(x)\varphi _{\lambda}.\end{equation*}
Hence we get
\begin{align}
&|\int (uv-u_xv_x) m_{x}(m\varphi _{\lambda})^{2q-1}\varphi _{\lambda} dx|
\leq CK^2\|m\varphi _{\lambda}\|_{2q}^{2q},
\end{align}
Note that $u,v,u_x,v_x,m,n,M \in L^{\infty}$. We achieve the following estimates
\begin{equation}
|-\frac{1}{2}\int M(m\varphi _{\lambda})^{2q-1}m\varphi _{\lambda} dx|\leq CK^2 \|m\varphi _{\lambda}\|^{2q}_{2q},
\end{equation}
and
\begin{equation}
|-\frac{1}{2}\int (uv-u_xv_x)(m\varphi _{\lambda})^{2q-1}m\varphi _{\lambda} dx|\leq CK^2 \|m\varphi _{\lambda}\|^{2q}_{2q}.
\end{equation}

From (3.32)-(3.34), this implies
\begin{equation}
\frac{d}{dt}\|m\varphi _{\lambda}\|_{2q}\leq CK^2\|m\varphi_{\lambda}\|_{2q}.
\end{equation}
By Gronwall's inequality, (3.35) implies the following estimate
\begin{equation}
\|m\varphi_{\lambda}\|_{2q}\leq \|m_0\varphi _{\lambda}\|_{2q}e^{CK^2t}.
\end{equation}
As the process of the estimation to (3.36), we deal with system (3.1)$_2$ is given by
\begin{equation}
\|n\varphi_{\lambda}\|_{2q}\leq \|n_0\varphi _{\lambda}\|_{2q}e^{CK^2t}.
\end{equation}
Add up (3.36) with (3.37), then by the Gronwall inequality yields that
\begin{equation}
(\|m\varphi_{\lambda}\|_{2q}+\|n\varphi_{2q}\|_{\infty})\leq (\|m_0\varphi_{\lambda}\|_{2q}+\|n_0\varphi _{2q}\|_{\infty})e^{CK^2t}.
\end{equation}
By virtue of the assumption (3.26), it follows that (3.27).

In view of the assumption (3.28) to obtain
\begin{equation*}
(m_0,n_0)\thicksim O(e^{-(1+\lambda)|x|}),~as~|x|\rightarrow \infty
\end{equation*}
Letting $q \rightarrow \infty$ in (3.36) and (3.37) and combing the above relation, we get
\begin{equation}
\|m\varphi_{\lambda}\|_{\infty}\leq \|m_0\varphi _{\lambda}\|_{\infty}e^{CK^2t},
\end{equation}
and
\begin{equation}
\|n\varphi_{\lambda}\|_{\infty}\leq \|n_0\varphi _{\lambda}\|_{\infty}e^{CK^2t}.
\end{equation}
Add up (3.39) with (3.40), then by the Gronwall inequality yields that
\begin{equation}
(\|m\varphi_{\lambda}\|_{\infty}+\|n\varphi_{\lambda}\|_{\infty})\leq (\|m_0\varphi_{\lambda}\|_{\infty}+\|n_0\varphi _{\lambda}\|_{\infty})e^{CK^2t}.
\end{equation}
On the other hand, by virtue of (3.28) and Theorem 2.2, we
deduce the last part of the theorem.
\end{proof}
\begin{remark3}
As long as the solution $z(t, x)$ exists, the result of Theorem
3.1 tells us that the solutions $(z,z_x)$ decay as $e^{-\theta|x|}$ when $|x|\rightarrow \infty$ for $\theta\in(0, 1).$ However, the momemtum $(m, n)$ can decay as
$e^{-(1+\lambda)|x|}$ as $|x|\rightarrow \infty$ for $\lambda \in (0,\infty).$
\end{remark3}

\section{Compactly supported initial data}
\newtheorem{theorem4}{Theorem}[section]
\newtheorem{lemma4}{Lemma}[section]
\newtheorem{corollary4}{Corollary}[section]
{\theoremstyle{definition}\newtheorem {remark4}{Remark}[section]}
\par
In this section, we reflect on the property of unique continuation which we have just shown the
Cauchy problem for the system (3.1) to exhibit. In the case of compactly supported
initial data unique continuation is essentially infinite speed of propagation of its support. Therefore, it is natural to ask the question: How will strong solutions behave at infinity when given compactly
supported initial data? We will need two ingredients in order to provide a sufficient answer.

Given initial data $z_0 \in H^{s}\times
H^{s}, s\geqslant 3,$ Theorem 2.1 ensures the local well-posedness
of strong solutions. Consider the following initial value problem
\begin{equation}
\left\{\begin{array}{ll}q_{t} = (uv-u_xv_x)(t,q) ,&t \in [0,T),\,x\in \mathbb{R}, \\
q(0,x) = x, &x\in \mathbb{R},\end{array}\right.
\end{equation}
where $u,v$ denotes the two component of solution $z$ to Eq.(3.1).
Since $z(t,.)\in H^3\times H^3 \subset C^m\times C^m$ with $0\leq m \leq \frac{5}{2},$
thus $z=(u,v) \in C^{1}([0,T)\times \mathbb{R},\mathbb{R}),$ applying the
classical results in the theory of ordinary differential equations,
one can obtain the following results of $q$ which is the key in the
proof of unique continuation of strong solutions to Eq.(4.1).

We now present the following two lemmas for our goal.
\begin{lemma4}\cite{Yan} Let $z_{0}\in H^{s}\times H^{s}$, $s \geq 2$. Then Eq.(4.1) has a
unique solution $q \in C^{1}([0,T)\times\mathbb{R},\mathbb{R}).$
Moreover, the map $q(t,\cdot)$ is an increasing diffeomorphism of
$\mathbb{R}$ with
$$
q_{x}(t,x)=\exp\left(\int_{0}^{t}(u_xn+v_{x}m)(s,q(s,x))ds\right)>0,
~~~(t,x)\in[0,T)\times\mathbb{R}.
$$
\end{lemma4}
\par
\begin{lemma4}\cite{Yan}
Let $z_{0}\in H^{s}\times H^{s-1}$, $s \geq 2$ and $T>0$ be the
maximal existence time of corresponding solution $z$ to Eq.(3.1).
Then for all $(t,x)\in
[0,T)\times\mathbb{R}$ we have
\begin{eqnarray}
m(t,q(t,x))q_{x}(t,x) = m_{0}(x)\exp{\int_0^t(u_xn+v_{x}m)(\tau,q(\tau,x))d\tau}, \\
n(t,q(t,x))q_{x}(t,x) = n_{0}(x)\exp{\int_0^t(u_xn+v_{x}m)(\tau,q(\tau,x))d\tau}.
\end{eqnarray}
\end{lemma4}

Now, utilizing the new form for the system (4.1) and our family of difieomorphisms given
by Lemma 4.1, we may now determine the behavior of our solutions at infinity when given compactly
supported initial data. This is provided via the following theorem.
\begin{theorem4}
Let $z\in C[0,T)\times C[0,T),s> \frac{5}{2},$ be a nontrivial solution of (3.1), with maximal time
of existence $T > 0,$ which is initially compactly supported on an interval $[a, b].$ Then we have
\begin{equation}
u(t,x)=\left\{\begin{array}{ll}\frac{1}{2}E_+(t)e^{-x} ,&x>q(t,b), \\
\frac{1}{2}E_-(t)e^{x} ,&x<q(t,a),\\
\end{array}\right.
\end{equation}
\begin{equation}
v(t,x)=\left\{\begin{array}{ll}\frac{1}{2}F_+(t)e^{-x} ,&x>q(t,b), \\
\frac{1}{2}F_-(t)e^{x} ,&x<q(t,a),\\
\end{array}\right.
\end{equation}
with $E_+(t):=\int_{q(t,a)}^{q(t,b)}e^y m(t,y)dy$, $E_-(t):=\int_{q(t,a)}^{q(t,b)}e^{-y} m(t,y)dy$,
$F_+(t):=\break\int_{q(t,a)}^{q(t,b)}e^y n(t,y)dy$ and $F_-(t):=\int_{q(t,a)}^{q(t,b)}e^{-y} n(t,y)dy.$  Moreover, $E_+(t)$, $E_-(t)$, $F_+(t)$ and $F_-(t)$ are continous non-vanishing functions with $E_+(0) = E_-(0)=F_+(0)=F_-(0)= 0$ and if $m_0$ and $n_0$ are non-negative, then $E_+, F_+$ strictly
increasing and $E_-, F_-$ strictly decreasing for $t\in [0, T).$
\end{theorem4}

Theorem 4.1 tells us that as long as the solution $z(x, t)$ exists, then it is positive at infinity and negative at negative infinity. We now proceed to the proof of the above result.

\begin{proof}
If $u_0$ and $v_0$ are initially supported on the compact interval $[a,b]$ then so are $m_0$ and$ n_0$. And from (4.2) and (4.3) it follows
that $m( t,\cdot),n(t,\cdot)$ is compactly supported with its support contained in the interval $[q(t,a), q(t,b)].$ We now use the relation $u = \frac{1}{2} e^{-|x|} \ast m$ and $v = \frac{1}{2} e^{-|x|} \ast n$ to write
\begin{align}
u(t,x)=\frac{e^x}{2}\int_{-\infty}^x e^y m(t,y)dy +\frac{e^x}{2}\int_{x}^{\infty} e^{-y} m(t,y)dy,\\
u_x(t,x)=-\frac{e^{-x}}{2}\int_{-\infty}^x e^y m(t,y)dy +\frac{e^x}{2}\int_{x}^{\infty} e^{-y} m(t,y)dy,
\end{align}
and
\begin{align}
v(t,x)=\frac{e^x}{2}\int_{-\infty}^x e^y n(t,y)dy +\frac{e^x}{2}\int_{x}^{\infty} e^{-y} n(t,y)dy,\\
v_x(t,x)=-\frac{e^{-x}}{2}\int_{-\infty}^x e^y n(t,y)dy +\frac{e^x}{2}\int_{x}^{\infty} e^{-y}, n(t,y)dy.
\end{align}
Assume that $m_0$ and $n_0$ are non-negative, then we obtain
\begin{align*}
u(t,x)+u_x(t,x)=\frac{e^x}{2}\int^{\infty}_x e^y m(t,y)dy\geq 0, \\
u(t,x)-u_x(t,x)=\frac{e^{-x}}{2}\int_{-\infty}^x e^y m(t,y)dy\geq 0,\\
v(t,x)+v_x(t,x)=\frac{e^x}{2}\int^{\infty}_x e^y n(t,y)dy\geq 0, \\
v(t,x)-v_x(t,x)=\frac{e^{-x}}{2}\int_{-\infty}^x e^y n(t,y)dy\geq 0.
\end{align*}
i.e. $|u_x|\leq u$ and $|v_x|\leq v$.
and then we define our functions
\begin{align*}
E_+(t)=\int_{q(t,a)}^{q(t,b)}e^y m(t,y)dy,~~E_-(t)=\int_{q(t,a)}^{q(t,b)}e^{-y} m(t,y)dy,\\
F_+(t)=\int_{q(t,a)}^{q(t,b)}e^y n(t,y)dy,~~F_-(t)=\int_{q(t,a)}^{q(t,b)}e^{-y} n(t,y)dy.
\end{align*}
we have that
\begin{align}
\nonumber u(t,x)=\frac{e^{-x}}{2}E_+(t),~~x>q(t,b), \\
\nonumber u(t,x)=\frac{e^{x}}{2}E_-(t),~~x<q(t,a),\\
\nonumber v(t,x)=\frac{e^{-x}}{2}F_+(t),~~x>q(t,b), \\
v(t,x)=\frac{e^{x}}{2}F_-(t),~~x<q(t,a),
\end{align}
therefore from differentiating (4.10) directly we get
\begin{align}
\nonumber \frac{e^{-x}}{2}E_+(t)=u(t,x)=-u_x(t,x)=u_{xx}(t,x),~~x>q(t,b), \\
\nonumber \frac{e^{x}}{2}E_-(t)=u(t,x)=u_x(t,x)=u_{xx}(t,x),~~x<q(t,a),\\
\nonumber \frac{e^{-x}}{2}F_+(t)=v(t,x)=-v_x(t,x)=v_{xx}(t,x),~~x>q(t,b), \\
\frac{e^{x}}{2}F_-(t)=v(t,x)=v_x(t,x)=v_{xx}(t,x),~~x<q(t,a).
\end{align}
Since $u(0,\cdot)$ and $v(0,\cdot)$ is supported in the interval $[a, b]$ this immediately gives us $E_+(0) = E_-(0) = 0$ and $F_+(0) = F_-(0) = 0$.

Since $m(t,\cdot)$ is supported in the interval $[q(t,a), q(t, b)]$, for each fixed $t$ we have
\begin{align}
 \frac{dE_+(t)}{dt}=\int_{q(t,a)}^{q(t,b)}e^y m_t(t,y)dy=\int_{-\infty}^{\infty}e^y m_t(t,y)dy.
\end{align}
Thus, we have
\begin{align*}
 &\frac{dE_+(t)}{dt}=\int_{q(t,a)}^{q(t,b)}e^y m_t(t,y)dy\\
 &=\int_{-\infty}^{\infty}e^y m_t(t,y)dy\\
 &=-\int_{-\infty}^{\infty}\frac{1}{2}[(uv-u_yv_y)m]_ye^ydy
 +\int_{-\infty}^{\infty}\frac{1}{2}(uv_y-vu_y)me^ydy\\
 &=\int_{-\infty}^{\infty}\frac{1}{2}(u-u_y)(v+v_y)me^ydy\geq 0.
\end{align*}
Nevertheless,
\begin{align*}
 &\frac{dE_-(t)}{dt}=\int_{q(t,a)}^{q(t,b)}e^{-y} m_t(t,y)dy\\
 &=\int_{-\infty}^{\infty}e^{-y} m_t(t,y)dy\\
 &=-\int_{-\infty}^{\infty}\frac{1}{2}[(uv-u_yv_y)m]_ye^{-y}dy
 +\int_{-\infty}^{\infty}\frac{1}{2}(uv_y-vu_y)me^{-y}dy\\
 &=\int_{-\infty}^{\infty}\frac{1}{2}(u+u_y)(v_y-v)me^{-y}ydy\leq 0,
\end{align*}
where the strict positivity of the relation above follows from our assumption that the solution is
nontrivial. Using the similar process gives the properties of $F_+$ and $F_-$.
This concludes the proof of Theorem 4.1.
\end{proof}

\section*{Acknowledgments} 
This work  was partially supported by the National Natural Science Foundation of China (Nos. 11401223, 11171295 and 61328103), NSF of Guangdong (No. 2015A030313424) and China Scholarship Council, and Qiao also thanks the Haitian Scholar Plan of Dalian University of Technology, the China state administration of foreign experts affairs system under the affiliation of China University of Mining and Technology, and the U.S. Department of Education GAANN project (P200A120256) for their cooperation in conducting the research program.

\medskip
Received April 2015; revised May 2015.
\medskip

\end{document}